\documentclass[pre,aps,reprint,twocolumn,showpacs]{revtex4}
\usepackage{graphicx}
\usepackage{amsmath}
\usepackage{amssymb}
\usepackage[latin2]{inputenc}
\def\slaninafigdir{.}
\begin{document}
\title{%
Inertial hydrodynamic ratchet: Rectification of colloidal flow in tubes of variable diameter 
}
\author{%
Franti\v{s}ek Slanina
}
\affiliation{%
Institute of Physics,
 Academy of Sciences of the Czech Republic,
 Na~Slovance~2, CZ-18221~Praha,
Czech Republic
}
\email{
slanina@fzu.cz
}
\begin{abstract}
We investigate analytically a microfluidic device consisting of a tube with 
non-uniform but spatially periodic diameter, where a fluid 
 driven back and forth by a  pump carries colloidal
 particles. Although the net flow of the   
fluid is zero, the particles move preferentially in one
direction due to ratchet mechanism, which occurs by simultaneous 
effect of inertial hydrodynamics and Brownian motion. We show that the
average current is strongly sensitive to particle size, thus
facilitating colloidal particle sorting.
\end{abstract}
%
%
%
%
%
%
%
\maketitle%
\section{Introduction}
Transport of soft matter in micropores \cite{mat_mul_03} and
nanopores \cite{huber_15} becomes increasingly important research
topic as fabrication of micro- and nanomachinery started to be widely
available. Among various applications let us mention for example
microfluidic 
lab-on-the-chip devices \cite{whitesides_06} or medical
microdiagnostics \cite{bha_bow_hou_tan_han_lim_10}. 

One of the key tasks such devices are expected to execute is sorting
of particles immersed in a fluid according to their size, shape,
elasticity and other physical 
properties \cite{saj_sen_14,xua_lee_14}. They may be micron-sized
colloidal particles, blood cells, 
microdroplets, bacteria etc. Generically, the fluid is let to flow
through a two-dimensional or quasi-one-dimensional chambers. In
two-dimensional case, the fluid passes through specially designed
two-dimensional system of obstacles, as 
in deterministic lateral displacement
devices \cite{hua_cox_aus_stu_04}, or flows through 
optical \cite{mcd_spa_dho_03} or acoustical \cite{pet_abe_swa_lau_07} latices. 

In quasi-one-dimensional structures
particles flow through tubes or channels of various shapes. 
Sorting can be achieved by pure hydrodynamic inertial effects, as
observed originally in experiments by Segr\'e and Silberberg \cite{seg_sil_62}.
 This idea led to
a great number of practical realizations in the last
years 
\cite{dic_iri_tom_ton_07,bha_kun_pap_08,dic_edd_hum_sto_ton_09,sun_liu_li_wan_xia_hu_jia_13,mar_ton_13,zho_pap_13,ami_lee_dic_14}. 
The main lesson from all these studies is, that curved 
shapes, either in the form of meanders or spirals, or in the form of
periodically varying diameter, greatly enhances the inertial effects
and thus the sorting capability of the device.

Alternatively, we can rely on the idea of Brownian
motors \cite{han_mar_08,reimann_02}. The motion of particles immersed in a fluid
 is rectified into a ratchet flow due to combination of 
Brownian motion, asymmetric entropic barriers caused by spatially periodic
variation of the tube, and periodic unbiased
external driving. Separation capabilities of such microdevices were
clearly demonstrated \cite{mat_mul_03,mar_bug_tal_sil_02,reg_bur_sch_rub_han_12}. In real
applications, hydrodynamics and Brownian motion 
always act together, and their interplay leads to 
 new phenomena, e. g. the hydrodynamically enforced entropic
trapping \cite{mar_str_sch_sch_han_13,mar_sch_str_sch_han_13}.

In this work we want to make a step in yet another direction, namely
toward the
combination of hydrodynamic inertial effects and Brownian motion. 
The original motivation for our work originates from the
experiment \cite{mat_mul_03} which was modeled theoretically
in \cite{ket_rei_han_mul_00} and reexamined recently \cite{mat_mul_gos_11}. Although
in \cite{mat_mul_gos_11} the experimentalists question their own original
interpretations, the setup used remains a paradigmatic one and
deserves attention. The rectification by purely hydrodynamic mechanism
in the same geometry was already demonstrated in numerical
simulations \cite{cis_vas_par_and_11}. 

Indeed, if the flow of the fluid  obeys
Stokes equation, the movement is perfectly reversible, thus leaving
the entropic barriers the only symmetry-breaking source of the ratchet
flow. In this article we show that inertial
hydrodynamic effects provide another symmetry-breaking ingredient,
which is sometimes dominant compared to the entropic
barriers and potentially even more efficient in terms of particle sorting.

To pursue this program it is first necessary to solve the flow in a
tube with variable diameter. Leaving aside the brute-force numerical
methods, this problem was already approached using perturbation
expansions \cite{cho_sod_72}, but mostly for slow-variation
expansion \cite{vandyke_87,kotorynski_95,sis_jin_zim_01}, which we consider insufficient for our 
purposes; or in the Stokes regime \cite{kit_dyk_97,mal_mit_adl_06}, thus
excluding the inertial effects from the beginning.
This is
also the approach of the
articles \cite{ket_rei_han_mul_00,mar_str_sch_sch_han_13,mar_sch_str_sch_han_13},
otherwise very closely related 
to our work. So, we would like to obtain more satisfactory, however approximate,
analytical solutions of the full Navier-Stokes equations. With a
solution at hand, we shall proceed with insertion of colloidal
particles. Such a two-step procedure is reflected by two Reynolds
numbers fixing the scales.  First, there is the
tube Reynolds number $\mathrm{Re}_t=Ud/\nu$, where $d$ is the average tube diameter
and $U$ is the average
velocity within the tube, defined through the volumetric flow $Q$ as
$U=4Q/\pi d^2$. Next, there is the particle Reynolds number
$\mathrm{Re}_p=UR^2/d\nu=(R/d)^2\,\mathrm{Re}_t$, where $R$ is the
particle radius. We shall work with tube Reynolds number $\mathrm{Re}_t\gtrsim 1$,
so that inertial effects are important, but the flow is still safely
stable laminar. On the other hand, for small enough particles we can
suppose 
$\mathrm{Re}_p\ll 1$, so that the perturbation caused by the particle
is described by Stokes equation. To be more specific, let us take as
an example the sorting apparatus investigated in
\cite{bha_kun_pap_08}. The authors used PDMS channels whose effective
diameter varied from 
$20\;\mu$m to $50\;\mu$m, the polystyrene spheres
had $2R=1.9\;\mu$m and the tube Reynolds number varied from $1$ to
$40$. This is about the scale we want to work with.

\begin{figure}[t]
\includegraphics[scale=0.3]{%
\slaninafigdir/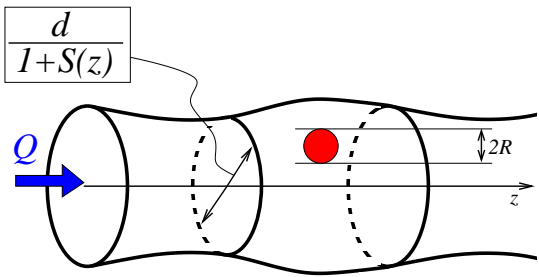}
\caption{%
Schematic sketch of the flow carrying a spherical particle with radius $R$ 
within a tube of varying diameter. The total volumetric flow of the
fluid is $Q$.}
\label{fig:scheme}
\end{figure}

\section{Flow in wavy tube}

The situation is shown schematically in Fig. \ref{fig:scheme}. The
coordinate axis $z$ coincides with the tube axis and the diameter of
the tube is governed by the periodic function $S(z)$. 
As the geometry of the tube is axially symmetric, we can use the
cylindrical coordinates and describe the 
flow by the (Stokes) stream function $\psi(\rho,z)$
satisfying a fourth-order partial differential equation
(see Appendix A). 

When the diameter of the tube  depends on the coordinate $z$
along the axis, it is convenient to change  further to
the generalized cylindrical coordinates 
\begin{equation}
\begin{split}
&\tilde{\rho}=(1+S(z))\rho\\
&\tilde{z}=z\;.
\end{split}
\label{eq:gencylindrical}
\end{equation}
 With such coordinates the tube wall is at fixed value
$\tilde{\rho}=d/2$, which simplifies the treatment of boundary
conditions. The function $S(z)$ describes variation of the diameter
around its medium value $d$ and it is supposed to vary periodically
along the tube. Therefore, we can expand it in terms of Fourier components
\begin{equation}
S(z)=\sum_{k=1,2,\ldots}\left(A_k\sin k\Omega z+B_k\cos k\Omega
z\right)\;.
\label{eq:fourierexpansionS}
\end{equation}

If $S=0$, inertial terms in the NS equations vanish. Therefore, our
strategy will be to expand in powers of the Fourier amplitudes of $S(z)$. The
stream function is then written as
\begin{equation}
\psi(\tilde{\rho},\tilde{z})=
\psi_0(\tilde{\rho})
+\psi_1(\tilde{\rho},\tilde{z})
+\psi_2(\tilde{\rho},\tilde{z})+\ldots
\end{equation}
where $\psi_n$ contains $n$-th powers of the amplitudes $A_k$,
$B_k$. The lowest inertial corrections are contained in
the term 
$\psi_1$. 

The equation for $\psi$ is non-linear, but we can transform it into a
chain of linear equations. Indeed, the equation for $\psi_0$ is linear
and if we already know functions $\psi_0$ through $\psi_{n-1}$, we can
insert them into an equation which is linear in the unknown
$\psi_n$. Here we shall stop at the lowest correction $\psi_1$.

Finding the zeroth term  $\psi_0$ is trivial as it does not depend on
$\tilde{z}$. The solution satisfying the proper boundary conditions is 
\begin{equation}
\psi_0(\tilde{\rho})=
\frac{4Q}{\pi}\left(\frac{\tilde{\rho}}{d}\right)^2
\left(1-2\left(\frac{\tilde{\rho}}{d}\right)^2\right)
\end{equation}
and we can see that the formula is
formally identical to the standard Poiseuille flow, but expressed in
the variable $\tilde{\rho}$.

Knowing $\psi_0$, we can write a linear equation for $\psi_1$. 
In analogy to (\ref{eq:fourierexpansionS}), we can expand it into sum
of Fourier components
\begin{equation}
\psi_1(\tilde{\rho},\tilde{z})=
\sum_{k=1,2,\ldots}\Big(\alpha_k(\tilde{\rho})\sin k\Omega \tilde{z}
+\beta_k(\tilde{\rho})\cos k\Omega
\tilde{z}\Big)\;.
\label{eq:fourierexpansionpsi}
\end{equation}
The boundary conditions at the tube wall and at the axis require that
\begin{equation}
\begin{split}
 \alpha_k\Big(\frac{d}{2}\Big)=
\alpha'_k\Big(\frac{d}{2}\Big)=
\beta_k\Big(\frac{d}{2}\Big)=
\beta'_k\Big(\frac{d}{2}\Big)=0
\\
 \alpha_k(0)=
\alpha'_k(0)=
  \beta_k(0)=
 \beta'_k(0)=0\;.
\end{split}
\end{equation}
This has an important consequence that the volumetric flow through the
tube resulting from $\psi_1$ is zero, so the total volumetric flow is
always $Q$ as given by $\psi_0$. Of course, the quantity which is
affected by non-zero $\psi_1$ is the pressure.

It turns out that the components with different index $k$ are
independent. This greatly simplifies the solution which at the end can
be written in a compact form as 
\begin{equation}
\begin{split}
\psi_1(\tilde{\rho},\tilde{z})=&
-\frac{2Q}{\pi}\,\mathrm{Im}\sum_{k}\Big[
g\Big(k\Omega\tilde{\rho};k\Omega\frac{d}{2},\frac{2k\Omega Q}{\pi\nu}\Big)\times\\
&\times(A_k-\mathrm{i}B_k)\,\exp(-\mathrm{i}k\Omega\tilde{z})\Big]\;.
\end{split}
\label{eq:solutionpsi}
\end{equation}
The complex function $g(x;r,t)$ of variable $x$ depends on parameters $r$ and
$t$ and can be found as a solution of the equation
\begin{equation}
(L_2-\mathrm{i}t\,L_1)g(x;r,t)=-K_2+\mathrm{i}t\,K_1
\label{eq:eqforg}
\end{equation}
where the operators $L_1$ and $L_2$ and functions $K_1$ and $K_2$ are
\begin{equation}
L_1=\Big(\frac{x}{r}\Big)^2\Big[1-\Big(\frac{x}{r}\Big)^2\Big]
\Big(-x^2\frac{d^2}{dx^2}+x\frac{d}{dx}+x^2\Big)
\end{equation}
\begin{equation}
\begin{split}
L_2=&
x^4\frac{d^4}{dx^4}
-2x^3\frac{d^3}{dx^3}
+\big(3x^2-2x^4\big)\frac{d^2}{dx^2}+\\
&+\big(-3x+2x^3\big)\frac{d}{dx}
+x^4
\end{split}
\end{equation}
\begin{equation}
K_1=\frac{x^6}{r^4}\Big[
\frac{8}{r^2}+1-\Big(\frac{8}{r^2}+2\Big)\Big(\frac{x}{r}\Big)^2
+\Big(\frac{x}{r}\Big)^4\Big]
\end{equation}
\begin{equation}
K_2=\frac{x^6}{r^2}\Big[
\frac{16}{r^2}+1-\Big(\frac{x}{r}\Big)^2\Big]\;.
\end{equation}

For certain specific values of $t$ it is possible to write the
solution of  (\ref{eq:eqforg}) in terms of 
hypergeometric functions. In general case we expand the function
$g(x;r,t)$ in powers of the parameter $t$ and solve separately the
equations for the expansion coefficients. So, if
$g(x;r,t)=\sum_{m=0}^\infty (\mathrm{i}t)^m\,g^{(m)}(x;r)$, we have
the chain of equations for the components $g^{(m)}(x;r)$
\begin{equation}
\begin{split}
L_2\,g^{(0)}&=-K_2\\
L_2\,g^{(1)}&=L_1\,g^{(0)}+K_1\\
L_2\,g^{(2)}&=L_1\,g^{(1)}\\
&\vdots
\end{split}
\label{eq:chainforg}
\end{equation}
which can be solved step by step in terms of Bessel functions. We defer
 explicit formulas for the solution to the Appendix B. 
Here we show only the series expansion in powers of $x$, where
the coefficients are finite sums of powers of the fraction $\xi=(x/r)^2$
\begin{equation}
\begin{split}
&g^{(1)}(x;r)=
-\frac{\xi^4}{144}
+\frac{\xi^3}{24}
-\frac{\xi^2}{16}
+\frac{\xi}{36}-\\
&-\Big(
\frac{\xi^4}{2304}
-\frac{5\,\xi^3}{1728}
+\frac{\xi^2}{144}
-\frac{\xi}{144}
+\frac{17}{6912}
\Big)x^2+\\
&+\ldots
\end{split}
\label{eq:expansiong1}
\end{equation}
which will be useful later.

In fact, the
 expansion parameter $t$ is proportional to the tube Reynolds number
 multiplied by the quantity $k\Omega d$. Therefore, the expansion can
 be considered as small-Reynolds number expansion. However, this holds
 only as long as $k\Omega$  is not too large. If the spatial frequency
 of the tube modulation $\Omega$ is large and/or if the modulation is
 not smooth but exhibits
 sharp edges (i. e. large $k$ must be taken into account) the
 expansion is no more useful and full solution of (\ref{eq:eqforg}) is
 necessary. 

We shall use the following specific form of diameter modulation
\begin{equation}
S(z)=A\sin 2\Omega z+B\cos\Omega z
\label{eq:profile}
\end{equation}
which is indeed fairly smooth. 
For small $\Omega d$ and small $\mathrm{Re}_t$, taking only lowest
terms in the $t$-expansion is 
a sensible approximation. At this point it is perhaps appropriate to
make a general remark concerning approximations made. As always in a
hydrodynamics problem it is always a Reynolds number which decides on
applicability of this or that approximation. In a complex geometry,
there are always several Reynolds numbers, and in our work we already
mentioned two of them, namely the tube and particle Reynolds
numbers. However, also the value $2k\Omega Q/\pi\nu$ assigned to the expansion parameter
$t$ can be regarded as a Reynolds number relating the flow velocity
$U$ to two geometric parameters, the average tube diameter $d$ and the
spatial period of the diameter variations $1/\Omega$. More precisely,
there are several harmonic components, i. e. terms with $k=1,2,3,\ldots$,
and each of them introduces its own length scale
$1/k\Omega$. Therefore, there is a Reynolds number for each of the
harmonic components. The small-$t$ expansion means that all these
Reynolds numbers must be $\ll 1$. For our target value of the tube Reynolds
number $\mathrm{Re}_t\simeq 2$ and supposing that the highest
harmonic component has $k=2$, as in (\ref{eq:profile}), this is
satisfied as long as $\Omega d\ll 1$, i. e. the spatial period is much
monger that the tube diameter. As a practical example, the experiments in
\cite{mat_mul_gos_11} have $d\simeq 3\;\mu$m, $\Omega\simeq
10^{5}\;\mathrm{m}^{-1}$ $\mathrm{Re}_t\simeq 0.2$ and therefore satisfy the condition. 
However, if the flow were faster,
e. g. $\mathrm{Re}_t\simeq 20$, the condition would be proportionally
stronger, i. e. spatial period would have been much larger than ten times the diameter.

\begin{figure}[t]
\includegraphics[scale=0.85]{%
\slaninafigdir/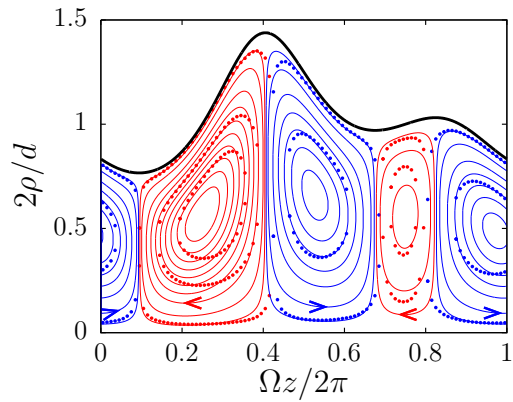}
\caption{%
Streamlines of the difference flow according to
(\ref{eq:solutiondeltapsi}) in a tube with profile given by
(\ref{eq:profile}) where $A=0.15$ and $B=0.2$. The orientation of the flow is
indicated by arrows. The thick curve indicates the tube wall. Using 
heavy dots we plot also the results of exact solution using OpenFoam, with
$\nu=10^{-6}\,\mathrm{m}^2\mathrm{s}^{-1}$ (water),
$d=0.3\,\mathrm{mm}$, $U=5\,\mathrm{mms}^{-1}$. (This means
$Q=0.245\ldots\,\mathrm{mm}^3\mathrm{s}^{-1}$, $\mathrm{Re}_t=1.5$.)
Note that the vertices would be totally absent if the flow were
described by Stokes equation.
}
\label{fig:streamlines}
\end{figure}

Rectification of the colloid flow occurs when the fluid is pumped
periodically back and forth by a piston. In such movement, the fluid
as a whole returns back to its original position after each period. We
shall consider the pumping adiabatic, composed of a first half-period of
stationary flow in one direction and a second half-period of
stationary flow in the opposite direction. The stream function in the
first and second half-period differ only in the sign of the
parameter $Q$. The difference flow is then simply 
the arithmetic average of the two. The stream function of the
difference flow, denoted $\delta\psi$, then contains only  even
powers of $Q$. Therefore, it is obtained from
(\ref{eq:solutionpsi}) by replacing $g(x;r,t)$ by expansion containing
only odd powers of $t$. If we keep only the lowest term, we get the
difference stream function
\begin{equation}
\begin{split}
\delta\psi&(\tilde{\rho},\tilde{z})=
\frac{4\Omega Q^2}{\pi^2\nu}\,\sum_{k}\Big[
k\,g^{(1)}\Big(k\Omega\tilde{\rho};\frac{1}{2}k\Omega d\Big)\times\\
&\times(B_k\sin k\Omega\tilde{z}-A_k\cos k\Omega\tilde{z})
\Big]+O(\mathrm{Re}_t^3)\;.
\end{split}
\label{eq:solutiondeltapsi}
\end{equation}
We show in Fig. \ref{fig:streamlines} streamlines of the flow
described by (\ref{eq:solutiondeltapsi}) in the tube of profile
(\ref{eq:profile}). We can see alternating clockwise and
counterclockwise vortexes, whose placing reflects the variations of
the tube diameter. The spatial extent of each vortex along the tube approximately
copies the segments of monotonous change of the diameter. Note that
the existence of the vortexes is purely inertial effect. Any solution
of the Stokes equation would give $\delta\psi$ identically zero. 

To check the quality of the two approximations made (i. e. small $S$
and small $t$), we solved the Navier-Stokes equations numerically,
using the package OpenFoam, for the parameters as shown in
Fig.  \ref{fig:streamlines}. If we compare the streamlines according
to  (\ref{eq:solutiondeltapsi}) with the exact result, we can see quite
good agreement. The position and shape of the vortexes is reproduced
well, the main difference being that the approximate streamlines
according to  (\ref{eq:solutiondeltapsi}) are more rounded. This indicates
that higher harmonic components, i. e. higher powers of $S$ would be
necessary for better agreement.

\section{Spherical particle in ambient flow}

Now let us insert a spherical particle of radius $R$ into the
flow.  The particle is considered neutrally-buoyant. The perturbation
to the ambient flow  
(\ref{eq:solutionpsi}) due to the presence of the particle and thus
the force acting on a spherical particle can be
computed by standard methods \cite{lan_lif_87,kim_kar_05}, expanding
the ambient flow into Taylor series around the center of the
sphere. In our actual calculations we took the Taylor series up to
quadratic terms only. Taking higher terms in this expansion would
result in terms of higher order in $R$ in the formula for particle
drift.
  
The perturbation is found by solving the Stokes equation,
 which is a valid approximation
if the particle size is much smaller than the tube diameter, so that 
$\mathrm{Re}_p\ll 1$ even if $\mathrm{Re}_t\gtrsim 1$. The
truncated Taylor expansion of the ambient flow
serves as boundary condition at
infinity. Certainly, this strategy fails when the distance from the
surface of the particle to the
wall is comparable with the particle diameter itself. In such case the
 hydrodynamic interactions are crucial and must be treated
 separately. However, when $d\gg R$, as we suppose throughout,
 the probability of particle being so close to the wall is very
 small. Therefore, we neglect this
effect here.

\begin{figure}[t]
\includegraphics[scale=0.85]{%
\slaninafigdir/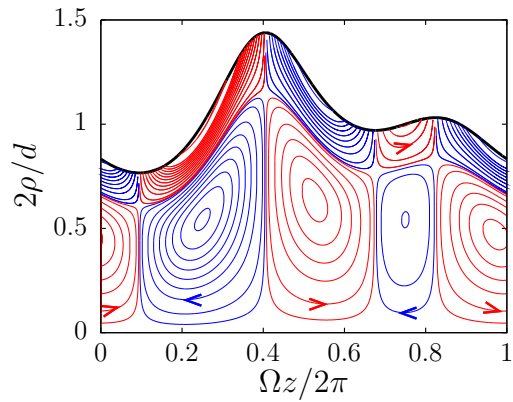}
\caption{%
Streamlines of the difference flow of particles, according to formula
(\ref{eq:solutiondeltapsipart}), in a tube with identical parameters
as in Fig. \ref{fig:streamlines}.
}
\label{fig:streamlines-part}
\end{figure}

The result valid for an axisymmetric flow is the following. 
If the ambient flow is
described by the stream function $\psi$, there is also a stream
function
\begin{equation}
\psi_\mathrm{p}=\psi+\frac{R^2}{6}\Big(
\frac{\partial^2}{\partial z^2} 
+\frac{\partial^2}{\partial \rho^2} 
-\frac{1}{\rho}\frac{\partial}{\partial\rho} 
\Big)\psi
+O\big(R^4\big)
\label{eq:streamfupart}
\end{equation}
  which corresponds to the
velocity field of the particle drift. We shall neglect terms of higher order
in the particle radius. In fact, they
are also of higher order in the particle Reynolds number. Thus we
arrive at nothing else than the Fax\'en law, which would be exact if
the ambient flow was a solution of Stokes, rather than Navier-Stokes
equation \cite{kim_kar_05}.

When the fluid is pumped back and forth, we can
establish the difference flow of the particles, in analogy with the
difference flow of the fluid. Denote $\delta\psi_{\mathrm{p}1}$ the
stream function for the difference 
flow of particles to the first order in $t$.  To this order we obtain

\begin{equation}
\begin{split}
\delta\psi_{\mathrm{p}1}(\tilde{\rho},\tilde{z})=&
\frac{4\Omega^3 Q^2}{\pi^2\nu}\,\sum_{k}\Big[k^3\Big(
g^{(1)\prime\prime}\big(k\Omega\tilde{\rho};\frac{k\Omega d}{2}\big)
-\\
-\frac{1}{k\Omega\tilde{\rho}}\,g^{(1)\prime}&\big(k\Omega\tilde{\rho};\frac{k\Omega d}{2}\big)
-g^{(1)}\big(k\Omega\tilde{\rho};\frac{k\Omega d}{2}\big)
\Big)
\times\\
&\times(B_k\sin k\Omega\tilde{z}-A_k\cos k\Omega\tilde{z})
\Big]
\end{split}
\label{eq:solutiondeltapsipart}
\end{equation}
where prime means differentiation with respect to the variable $x$. The
streamlines corresponding to particle drift according to
(\ref{eq:solutiondeltapsipart}) are  plotted in
Fig. \ref{fig:streamlines-part}. We can see marked difference from the
fluid streamlines shown in Fig. \ref{fig:streamlines}. There are
vortexes as in Fig. \ref{fig:streamlines}, but near the walls there
are also ``half-vortexes'' which imply that at some places the drift
pushes the particles towards the wall while at other places the
particles are pulled away. To understand this effect properly, we must
keep in mind that the difference flow depicts what happens after a whole
period of pumping is completed. During the period the particle can
follow some trajectory which may be complicated, but at the end of the
period  the particle is found shifted along the streamlines of
Fig.  \ref{fig:streamlines-part} from its initial position. The pushing and
pulling is therefore a summary effect of the movement over the whole period. 
Again, we should stress that the non-zero difference flow
(\ref{eq:solutiondeltapsipart}) is purely inertial effect and would be
exactly zero if the ambient flow were described by Stokes equation.
Finally, let us note that the hydrodynamic interaction of the particle
with the wall, which is not considered here, would lead to modification of
the half-vertexes on the scale $\simeq R$ from the wall. As $R\ll d$,
we neglect it here.

The presence of ``half-vortexes'' also means that the particle drift,
when integrated over the cross-section of the tube, depends on the
coordinate $z$. This suggests an approximate mapping of the particle
drift on an effective one-dimensional movement. The drift velocity
imposed on the particle in such mapping is simply, according to
the general properties of the stream function,
$w(z)=8\big(\delta\psi_{\mathrm{p}1}(d/2,z)-\delta\psi_{\mathrm{p}1}(0,z)\big)/d^2$. For
the specific profile (\ref{eq:profile}) we get
\begin{equation}
\begin{split}
w(z)=\frac{16\,Q^2R^2\Omega^3}{3\pi^2d^2\nu}
\Big[
&g^{(1)\prime\prime}\big(\frac{1}{2}\Omega d;\frac{1}{2}\Omega d\big)
B\sin \Omega z-\\
-8\,&g^{(1)\prime\prime}\big(\Omega d;\Omega d\big)
A\cos 2\Omega z
\Big]\;.
\end{split}
\label{eq:effdirft}
\end{equation}

\section{Ratchet effect}

The drift (\ref{eq:effdirft}) is a periodic function of $z$ and
therefore does not impose any ratchet current by itself. To see the
rectification of the particle flow, hydrodynamics must be accompanied
by diffusion. The full analysis of the hydrodynamic ratchet would require solving
the three-dimensional axially symmetric diffusion problem in the tube
with spatially dependent drift
given by (\ref{eq:solutiondeltapsipart}). However, here we remain on a
simpler level and estimate the ratchet effect by mapping on a
one-dimensional diffusion problem. In fact, we already started this
program by computing the effective drift (\ref{eq:effdirft}). 
Using the standard Fick-Jacobs
mapping \cite{zwanzig_92,reg_sch_bur_rub_rei_han_06}, we 
have, to first order in $S(z)$, the stationary diffusion equation
\begin{equation}
0=\frac{\partial}{\partial z}\Big[
(-w(z)\pm h+2DS'(z))\,P(z)\Big]
+D\frac{\partial^2}{\partial z^2}P(z)\;.
\label{eq:diffusiononedimensional}
\end{equation}
The diffusion
coefficient for a sphere is $D=kT/(6\pi\rho_f\nu R)$, where $\rho_f$
is here the density if the fluid.
The term $h$ accounts for steady driving due to the periodic
pumping. Its amplitude is $h=4Q/(\pi d^2)$ and the sign is positive in
one half-period and negative in the other one. In fact, this term
stands for all terms with even power of $Q$ in the expansion of the
ambient flow. As the leading term of this type is independent of $z$,
we can safely approximate it by constant $h$. 
The equation (\ref{eq:diffusiononedimensional}) describes a standard
Brownian motor which is exactly solved in the
literature \cite{reimann_02}. 

The advection term contains the hydrodynamic drift $w(z)$ and the term
$DS'(z)$ describing the entropic barrier. Using the average
velocity $U=4Q/\pi d^2$, the condition for the former
to dominate the
latter can be written as
\begin{equation}
\frac{2}{3}U^2R^3\,\rho_f\gg kT\;.
\label{eq:conditionhydrodymanic}
\end{equation}
 Considering
neutrally-buoyant colloidal particle, this condition can be
understood so that the kinetic energy of the particle as carried by
the fluid is much larger that the thermal quantum $kT$. For parameters
used in Fig. \ref{fig:streamlines} and at normal laboratory temperature the condition
(\ref{eq:conditionhydrodymanic}) is satisfied for particles larger
than about $0.5\mu\mathrm{m}$. Therefore, the typical particles used
in experiments like \cite{mat_mul_03} or all kinds of blood cells satisfy
(\ref{eq:conditionhydrodymanic}). 

\begin{figure}[t]
\includegraphics[scale=0.85]{%
\slaninafigdir/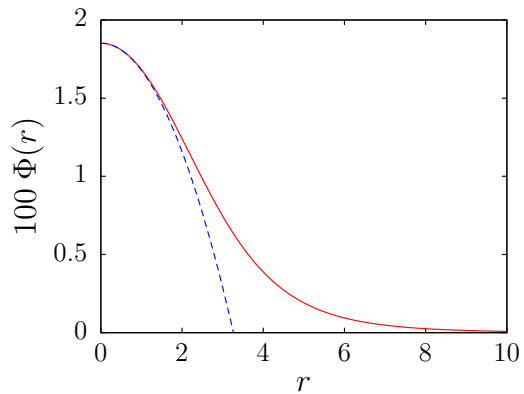}
\caption{%
Graph of the function $\Phi(r)$ defined by (\ref{eq:phi}). The
approximation (\ref{eq:phiexpansion}) is drawn by the dashed line
}
\label{fig:Phi}
\end{figure}

The exact solution of the motor described by
(\ref{eq:diffusiononedimensional}), as shown in \cite{reimann_02}, is
given by a complicated formula. If we are in the hydrodynamic regime
(\ref{eq:conditionhydrodymanic}), and if we expand the solution in
powers of  $A/kT$ and $B/kT$, and simultaneously in the
quantity $Q/(d^2\Omega kT)$, we obtain for the average
ratchet velocity of the particle
\begin{equation}
\langle v_\mathrm{p}\rangle=
\frac{2}{9}\Big(\frac{6\pi\rho_f}{kT}\Big)^4
\frac{\nu}{\Omega}\Big(\frac{4Q}{\pi d^2}\Big)^8\,R^{10}\,\Phi(\Omega d)\,A^2B
\label{eq:avervelocity}
\end{equation}
where
\begin{equation}
\Phi(r)=r^6\big(g^{(1)\prime\prime}(r;r)\big)^2\,g^{(1)\prime\prime}(r/2;r/2)\;.
\label{eq:phi}
\end{equation}
For tubes narrow with respect to the spatial period of the diameter
modulation we can use (\ref{eq:expansiong1}) and
expand the function
$\Phi(r)$ in powers 
\begin{equation}
\Phi(r)=\frac{1}{54}-\frac{r^2}{576}+O(r^4)\;.
\label{eq:phiexpansion}
\end{equation}
In the opposite limit $r\to\infty$ the function $\Phi(r)$
decreases to zero, reflecting the fact that the ratchet effect vanishes
if the diameter changes too fast along the tube. The full form of the
function $\Phi(r)$ is shown in Fig. \ref{fig:Phi}, together with the
approximation (\ref{eq:phiexpansion}), which is good as long as
$r\lesssim 2$. In practice this condition just tells us  that the spatial
period of tube modulation must be at least half of the tube diameter. 

We can see that the dependence of the ratchet current
on the particle diameter is extremely strong, much stronger than what
would be seen with entropic barriers only. This can be traced to
the diameter dependence of the hydrodynamic drift
(\ref{eq:effdirft}). We believe here is a potential for practicable
particle sorting. In this direction, it is essential to choose the
tube variation in an optimal way. Unfortunately, there is no obvious
way how to prescribe the optimal shape, as is the case already with
the simplest Brownian ratchet solved in \cite{reimann_02}. For
optimization purposes the limits of small amplitudes $A$ and $B$ used when
deriving Eq. (\ref{eq:avervelocity}) are not sufficient and numerical
evaluation of the exact formula (Equations (5.2-3) in
\cite{reimann_02}) is 
unavoidable.  Some
statements can be made, though. In our case, the effective ratchet
potential has its origin in the half-vortexes, as exemplified in Fig.
\ref{fig:streamlines-part}. These half-vortexes stem
from the second derivative of the ambient flow and therefore
it seems plausible to suppose that they are more pronounced if the
second derivative of the ambient flow is higher. This may happen if
the flow is less ``smooth'', e. g. for tubes with sharp
edges. However, this is just a hypothesis which would require thorough
testing.

\section{Conclusions}
To conclude, we provided an analytic solution of the problem of a
spherical colloidal particle carried by a fluid flow in a tube of
variable diameter.  The primary approximation is taking only lowest
term in the expansion in powers of the amplitude of diameter
modulation. The solution is approximate but agrees well
with direct numerical solution of NS equations in the regime of
Reynolds number we used, $\mathrm{Re}_t\simeq 1.5$. Further
improvement is possible taking into account higher powers of the
amplitude, but mixing together various harmonic modes would make the
analysis rather complicated.

The fluid is pumped back and forth so that the
average fluid flow is zero. Inertial
effects are non-negligible and lead to non-trivial drift acting on the
particle. Attention should be paid on the inertial effects when
interpreting the experimental results, as the regime in which the flow
oscillates with no average volumetric bias is not the same as regime in which the
pressure oscillates symmetrically with no pressure bias. With tube
profile breaking the mirror symmetry the inertial effects make the two
regimes different.

Combining the inertial hydrodynamic particle drift with diffusion, a ratchet
effect occurs, rectifying the average movement of  the particle. We
analyzed it by standard mapping on a one-dimensional problem. We
found a criterion when this inertial hydrodynamic effect dominates the effect of
entropic barriers (which is nevertheless always present).  We
showed that the  average ratchet velocity depends  very strongly on the
particle radius. The dependence is much more pronounced than with
entropic barriers. Therefore, inertial hydrodynamics  opens
a promising perspective for new approach in sorting 
particles according to their size, which may be more efficient than
sorting with entropic barriers only. Detailed comparison of the two
regimes, as well as the study of the crossover between the two, which
must occur when the fluid velocity is gradually increased, would
deserve special study, which we leave for future. 

Another interesting
ramification of this hydrodynamic study concerns the situation in
which Brownian particles in asymmetric channels are driven by
electrostatic or gravitational force. The interplay between these
external forces and hydrodynamics was already studied
\cite{mar_sch_str_sch_han_13,mar_str_sch_sch_han_13}, but as far as we
know inertial effects were left aside. But as a matter of fact driven particles immersed
in the fluid induce flows in the otherwise resting fluid and
 these
flows can have components which go beyond the Stokes
approximation. Existence of such inertial effects was demonstrated
in sedimentation experiments, e. g. in \cite{che_mcl_90}, but their
relevance for electrostatically driven transport in micropores remains
to be studied.

A natural question arises what are the limits of applicability of the
current analytic approximation. We checked, as also shown in
Fig. \ref{fig:streamlines}, that the approximation gives good results
when compared to exact numerical solution of NS equations for tube
Reynolds numbers up to about $\mathrm{Re}_t\simeq 1.5$. For example in
\cite{mat_mul_gos_11} the data suggest Reynolds numbers up to
$\mathrm{Re}_t\simeq 0.7$, so the present theory should be well
appropriate to describe their experiments. 
We believe that
our theory is in principle applicable to even higher Reynolds
numbers. Currently we do not have a systematic data available, but we
can estimate the limits from other works using the same level of
approximation in the study of Segr\'e-Silberberg effect. For example in
\cite{asmolov_02} it is claimed that their approximation remains valid
up to Reynolds numbers in the range of several hundreds. We
hypothesize that our approximation may be restricted by similar limits. 
In practice it means that e.g. the experiments with
sorting polystyrene spheres by Segr\'e-Silberberg effect
\cite{bha_kun_pap_08}, where  the Reynolds number achieved was $\mathrm{Re}_t\simeq
40$, fall probably within the validity of the present theory. Also the
experiment with tumor cell separation in
\cite{sun_liu_li_wan_xia_hu_jia_13} used Reynolds numbers at most
$100$, hence we believe our method would be applicable too. On the
other hand, in another tumor-cell sorting
experiment reported in \cite{mac_kim_ars_hur_dic_11}, the Reynolds
number was much higher,   $\mathrm{Re}_t\simeq 
1500$. It would require a separate study to see,  whether our theory
is still valid for such rapid flows.

Finally, we believe that the method used in this work can be adapted
to broader range of problems in quasi-one-dimensional transport. For
example, there is growing interest in the dynamics of active
particles in confined geometries
\cite{bec_dil_low_rei_vol_vol_16}. Separation of such particles is an
interesting problem \cite{ber_etal_13} and our method could be perhaps
adapted by simply replacing Eq. (\ref{eq:streamfupart}) by stream
function appropriate to active particles. Investigations in this
direction are in course. 

Another modification can be inspired by the recent work
\cite{ver_par_pos_15} where angular velocity component is forced onto
granular flow by spiral structures, not unlike rifling of gun
barrels. If such rifling were possible in a microfluidic channel, it
could also be used to induce inertial ratchet effect. In this case,
the system is still axially symmetric, if we consider the rifling
infinitesimally fine. Therefore 
our method is still applicable in principle, but must be generalized
by allowing non-zero angular velocity 
in the flow. On the better side, in such setup we could avoid the
(rather big) technical 
complications arising from the use of generalized cylindrical
coordinates. Other kinds of curved and spiraling shapes of the tubes
are also conceivable, but as soon as we loose the axial symmetry,
analytical approaches become too difficult.

\begin{acknowledgments}
I gladly acknowledge inspiring discussions with P. Chvosta and A. Ryabov. 

\end{acknowledgments}
\appendix

\section{Stokes stream function}
An axisymmetric flow with zero azimuthal component can be expressed in
terms of the Stokes stream function $\psi(\rho,z)$, related to the
cylindrical components of the velocity field as  
\begin{equation}
\begin{split}
u_\rho = -\frac{1}{\rho}\frac{\partial\psi}{\partial z}\\
u_z = \frac{1}{\rho}\frac{\partial\psi}{\partial \rho}\; .\\\;
\end{split}
\end{equation}
Then, the continuity equation is satisfied automatically and the
Navier-Stokes equation translates in the fourth-order  equation for
$\psi(\rho,z)$, which can be written as
\begin{widetext}
\begin{equation}
\begin{split}
\frac{1}{\rho^2}
\left(\frac{\partial\psi}{\partial z}\frac{\partial}{\partial\rho}-\frac{\partial\psi}{\partial\rho}\frac{\partial}{\partial z}\right)\left(
\frac{\partial^2\psi}{\partial\rho^2}
+\frac{\partial^2\psi}{\partial z^2}\right)
+
\frac{1}{\rho^3}\left(
\frac{\partial\psi}{\partial\rho}\frac{\partial^2\psi}{\partial\rho\partial z}
-3\frac{\partial\psi}{\partial z}\frac{\partial^2\psi}{\partial\rho^2}
-2\frac{\partial\psi}{\partial z}\frac{\partial^2\psi}{\partial z^2}
\right)
+
3\frac{1}{\rho^4}
\frac{\partial\psi}{\partial\rho}\frac{\partial\psi}{\partial z}
=\\
-\nu\left\{
\frac{1}{\rho}\left(
\frac{\partial^4\psi}{\partial\rho^4}
+2\frac{\partial^4\psi}{\partial\rho^2\partial z^2}
+\frac{\partial^4\psi}{\partial z^4}
\right)
-2
\frac{1}{\rho^2}\frac{\partial}{\partial\rho}\left(
\frac{\partial^2\psi}{\partial\rho^2}
+\frac{\partial^2\psi}{\partial z^2}
\right)
+
3\frac{1}{\rho^3}
\frac{\partial^2\psi}{\partial\rho^2}
-
3\frac{1}{\rho^4}
\frac{\partial\psi}{\partial\rho}
\right\}\;.
\end{split}
\label{eq:appendix-eqforpsi}
\end{equation}
\end{widetext}
The next step is writing the equation in terms of the generalized
cylindrical coordinates (\ref{eq:gencylindrical}). The derivatives transform as 
\begin{equation}
\begin{split}
&\frac{\partial}{\partial{\rho}}=\big(1+S(\tilde{z})\big)\,\frac{\partial}{\partial\tilde{\rho}}\\
&\frac{\partial}{\partial{z}}=\frac{\partial}{\partial\tilde{z}}+
\frac{S'(\tilde{z})}{1+S(\tilde{z})}\,\tilde{\rho}\,\frac{\partial}{\partial\tilde{\rho}}
\end{split}
\label{eq:appendix-gencylindricalderivatives}
\end{equation}
and inserting them into (\ref{eq:appendix-eqforpsi}) we obtain the
desired equation for the stream function. When we neglect in this
equation all terms of higher order than linear in the modulation amplitude, we
finally arrive at equations for $\psi_0(\tilde{\rho})$ and $\psi_1(\tilde{\rho},\tilde{z})$.

\section{Explicit formulas for expansion in powers of $t$}
The solution of the equation
\begin{equation}
(L_2-\mathrm{i}t\,L_1)g(x;r,t)=-K_2+\mathrm{i}t\,K_1
\label{eq:appendix-eqforg}
\end{equation}
is expanded as
$g(x;r,t)=\sum_{m=0}^\infty (\mathrm{i}t)^m\,g^{(m)}(x;r)$.
To establish the expansion coefficients $g^{(m)}$ we have to  solve
the chain of equations  (\ref{eq:chainforg}). 
Generally, the real functions $g^{(m)}$  can be written in terms of a
particular solution and a combination of independent 
solutions of the homogeneous equation $L_2\phi(x)=0$. 
As two independent solutions of the homogeneous equation, satisfying
the proper boundary conditions, we choose the functions 
\begin{equation}
H_1(x)=xI_1(x)
\end{equation}
\begin{equation}
\begin{split}
&H_2(x)=\\&x^2K_1(x)\Big(xI_0^2(x)-xI_1^2(x)-2I_0(x)I_1(x)\Big)+\\
&+2xI_1(x)\int_0^xx'K_1(x')I_1(x')\,dx'\;.\\   \;
\end{split}
\end{equation}
The other two linearly independent solutions do not satisfy the boundary conditions at the
axis of the tube and would be taken into account only if the tube
 contained a concentric hard core.

Then, the $m$-th term can be written as a linear combination of the
functions $H_1$ and $H_2$ plus the particular solution, i. e.  
\begin{equation}
\begin{split}
g^{(m)}(x;r)=\,&h_1^{(m)}(r)\,H_1(x)+h_2^{(m)}(r)\,H_2(x)+\\&+P^{(m)}(x;r)\;.
\end{split}
\end{equation}
We show here explicitly the results for $m=0$ and $m=1$. In the zeroth
order, the particular solution is  
\begin{equation}
P^{(0)}(x;r)=\Big(\frac{x}{r}\Big)^4-\Big(\frac{x}{r}\Big)^2
\end{equation}
and the coefficients in the linear combination are
\begin{equation}
h_1^{(0)}(r)=-2\Big(K_1(r)+
2I_1(r)\,\phi_2(r)\Big)\,\phi_3(r)
\end{equation}
\begin{equation}
\begin{split}
h_2^{(0)}(r)=2I_1(r)\,\phi_1(r)\,\phi_3(r)
\end{split}
\end{equation}
where we defined  auxiliary functions
\begin{equation}
\phi_1(r)=
\frac{1}{r^2I_0^2(r)-r^2I_1^2(r)-2rI_0(r)I_1(r)}
\end{equation}

\begin{equation}
\phi_2(r)=
\phi_1(r)\int_0^rx'K_1(x')I_1(x')\,dx'
\end{equation}
and
\begin{equation}
\phi_3(r)=\frac{1}{r^2
  \Big(K_1(r)I_0(r)+I_1(r)K_0(r)\Big)}\;.
\end{equation}

In the first order, the particular solution can be written as
\begin{equation}
\begin{split}
&P^{(1)}(x;r)=\\&2x\,\Big(I_1(x)f_1(x;r)-K_1(x)f_2(x;r)\Big)\,h_2^{(0)}(r)\\\;
\end{split}
\end{equation}
and the coefficients as
\begin{equation}
h_1^{(1)}(r)=-2\Big(
f_1(r;r)+
2\,f_2(r;r)\,\phi_2(r)
\Big)h_2^{(0)}(r)
\end{equation}
\begin{equation}
h_2^{(1)}(r)=2f_2(r;r)\,\phi_1(r)\,
h_2^{(0)}(r)
\end{equation}
where
\begin{equation}
\begin{split}
f_1(x;r)=\int_0^x\int_0^{x'}x'K_1(x')\Big(I_1(x')K_1(y)-\\-K_1(x')I_1(y)\Big)f_3(y;r)\,dy\,dx'
\end{split}
\end{equation}
\begin{equation}
\begin{split}
f_2(x;r)=\int_0^x\int_0^{x'}x'I_1(x')\Big(I_1(x')K_1(y)-\\-K_1(x')I_1(y)\Big)f_3(y;r)\,dy\,dx'
\end{split}
\end{equation}
and
\begin{equation}
\begin{split}
&f_3(y;r)=\Big[\Big(\frac{y}{r}\Big)^4-\Big(\frac{y}{r}\Big)^2\Big]\times\\&\times I_1(y)
\Big(K_1(y)I_0(y)+I_1(y)K_0(x)\Big) \;.
\end{split}
\end{equation}


\begin{thebibliography}{99}
\bibitem{mat_mul_03}
S. Matthias and F. M\"uller,
Nature
 {\bf 424},
 53
 (2003).

\bibitem{huber_15}
P. Huber,
J. Phys.: Condens. Matter
 {\bf 27},
 103102
 (2015).

\bibitem{whitesides_06}
G. M. Whitesides,
Nature
 {\bf 442},
 368
 (2006).

\bibitem{bha_bow_hou_tan_han_lim_10}
A. A. S. Bhagat, H. Bow, H. W. Hou, S. J. Tan, J. Han, and C. T. Lim,
Med. Biol. Eng. Comput.
 {\bf 48},
 999
 (2010).

\bibitem{saj_sen_14}
P. Sajeesh and A. K. Sen,
Microfluidics and Nanofluidics
 {\bf 17},
 1
 (2014).

\bibitem{xua_lee_14}
J. Xuan and   M. L. Lee,
Anal. Methods
 {\bf 6},
 27
 (2014).

\bibitem{hua_cox_aus_stu_04}
L. R. Huang, E. C. Cox, R. H. Austin, J. C. Sturm,
Science
 {\bf 304},
 987
 (2004).

\bibitem{mcd_spa_dho_03}
M. P. MacDonald, G. C. Spalding, and K. Dholakia,
Nature
 {\bf 426},
 421
 (2003).

\bibitem{pet_abe_swa_lau_07}
F. Petersson, L. {\AA}berg, A.-M. Sw\"ard-Nilsson, and T. Laurell,
Anal. Chem.
 {\bf 79},
 5117
 (2007).

\bibitem{seg_sil_62}
G.  Segr\'e and A.  Silberberg,
J. Fluid Mech.
 {\bf 14},
 136
 (1962).

\bibitem{dic_iri_tom_ton_07}
D. Di Carlo, D. Irimia, R. G. Tompkins, and M. Toner,
Proc. Nat. Acad. Sci. USA
 {\bf 104},
 18892
 (2007).

\bibitem{bha_kun_pap_08}
A. A. S. Bhagat, S. S. Kuntaegowdanahalli, and I. Papautsky,
Phys. Fluids
 {\bf 20},
 101702
 (2008).

\bibitem{dic_edd_hum_sto_ton_09}
D. Di Carlo, J. F. Edd, K. J. Humphry, H. A. Stone, and M. Toner,
Phys. Rev. Lett.
 {\bf 102},
 094503
 (2009).

\bibitem{sun_liu_li_wan_xia_hu_jia_13}
J. Sun, C. Liu, M. Li, J. Wang, Y. Xianyu, G. Hu and X. Jiang,
Biomicrofluidics
 {\bf 7},
 011802
 (2013).

\bibitem{mar_ton_13}
J. M. Martel and M. Toner,
Scientific Reports
 {\bf 3},
 3340
 (2013).

\bibitem{zho_pap_13}
J. Zhou and I. Papautsky,
Lab Chip
 {\bf 13},
 1121
 (2013).

\bibitem{ami_lee_dic_14}
H. Amini, W. Lee, and D. Di Carlo,
Lab Chip
 {\bf 14},
 2739
 (2014).

\bibitem{han_mar_08}
P. H\"anggi and F. Marchesoni,
Rev. Mod. Phys.
 {\bf 81},
 387
 (2009).

\bibitem{reimann_02}
P. Reimann,
Phys. Rep.
 {\bf 361},
 57
 (2002).

\bibitem{mar_bug_tal_sil_02}
C. Marquet, A. Buguin, L. Talini, and P. Silberzan,
Phys. Rev. Lett.
 {\bf 88},
 168301
 (2002).

\bibitem{reg_bur_sch_rub_han_12}
D. Reguera, A. Luque, P. S. Burada, G. Schmid, J. M. Rub\'{\i}, and P. H\"anggi,
Phys. Rev. Lett.
 {\bf 108},
 020604
 (2012).

\bibitem{mar_str_sch_sch_han_13}
S. Martens, A. V. Straube, G. Schmid, L. Schimansky-Geier, and P. H\"anggi,
Phys. Rev. Lett.
 {\bf 110},
 010601
 (2013).

\bibitem{mar_sch_str_sch_han_13}
S. Martens, G. Schmid, A. V. Straube, L. Schimansky-Geier, P. H\"anggi,
Eur. Phys. J. Spec. Topics
 {\bf 222},
 2453
 (2013).

\bibitem{ket_rei_han_mul_00}
C. Kettner, P. Reimann, P. H\"anggi, and F. M\"uller,
Phys. Rev. E
 {\bf 61},
 312
 (2000).

\bibitem{mat_mul_gos_11}
K. Mathwig, F. M\"uller, and U. G\"osele,
New J. Phys.
 {\bf 13},
 033038
 (2011).

\bibitem{cis_vas_par_and_11}
R. L. C. Cisne Jr., T. F. Vasconcelos, E. J. R. Parteli, and J. S. Andrade Jr.,
Microfluidics and  Nanofluidics
 {\bf 10},
 543
 (2011).

\bibitem{cho_sod_72}
J. C. F. Chow and K. Soda,
Phys. Fluids
 {\bf 15},
 1700
 (1972).

\bibitem{vandyke_87}
M. Van Dyke,
Adv. Appl. Mech.
 {\bf 25},
 1
 (1987).

\bibitem{kotorynski_95}
W. P. Kotorynski,
Computers and Fluids
 {\bf 24},
 685
 (1995).

\bibitem{sis_jin_zim_01}
S. Sisavath, X. Jing, and R. W. Zimmerman,
Phys. Fluids
 {\bf 13},
 2762
 (2001).

\bibitem{kit_dyk_97}
P. K. Kitanidis and B. B. Dykaar,
Transport in Porous Media
 {\bf 26},
 89
 (1997).

\bibitem{mal_mit_adl_06}
A. E. Malevich, V. V. Mityushev, and P. M. Adler,
Acta Mechanica
 {\bf 182},
 151
 (2006).

\bibitem{lan_lif_87}
L. D. Landau and E. M. Lifshitz,
{\it Fluid Mechanics}
 (Butterworth-Heinemann, Oxford, 1987).

\bibitem{kim_kar_05}
S. Kim and S. J. Karilla,
{\it Microhydrodynamics}
 (Dover Publications, New York, 2005).

\bibitem{zwanzig_92}
R. Zwanzig,
J. Phys. Chem.
 {\bf 96},
 3926
 (1992).

\bibitem{reg_sch_bur_rub_rei_han_06}
D. Reguera, G. Schmid, P. S. Burada, J. M. Rub\'\i, P. Reimann, and P. H\"anggi,
Phys. Rev. Lett.
 {\bf 96},
 130603
 (2006).

\bibitem{che_mcl_90}
P. Cherukat and J. B. McLaughlin,
Int. J. Multiphase Flow
 {\bf 16},
 899
 (1990).

\bibitem{asmolov_02}
E. S. Asmolov,
Phys. Fluids
 {\bf 14},
 15
 (2002).

\bibitem{mac_kim_ars_hur_dic_11}
A. J. Mach, J. H. Kim, A. Arshi,  S. C. Hur, and   D. Di Carlo,
Lab Chip
 {\bf 11},
 2827
 (2011).

\bibitem{bec_dil_low_rei_vol_vol_16}
C. Bechinger, R. Di Leonardo, H. L\"owen, C. Reichhardt, G. Volpe, and G. Volpe,
Rev. Mod. Phys. (accepted), arXiv:1602.00081.

\bibitem{ber_etal_13}
I. Berdakin, Y. Jeyaram, V. V. Moshchalkov, L. Venken, S. Dierckx, S. J. Vanderleyden, A. V. Silhanek, C. A. Condat, and V. I. Marconi,
Phys. Rev. E
 {\bf 87},
 052702
 (2013).

\bibitem{ver_par_pos_15}
F. Verb\"ucheln, E. J. R. Parteli, and  T. P\"oschel,
Soft Matter
 {\bf 11},
 4295
 (2015).

%
%
\end{thebibliography}
\end{document}